\documentclass[sigconf]{acmart}
\usepackage{multirow}
\usepackage{makecell}
\AtBeginDocument{%
  }

\begin{document}

\title{ReCode: Improving LLM-based Code Repair with Fine-Grained Retrieval-Augmented Generation}

\author{Yicong Zhao}

\affiliation{%
  \department{College of Computer Science and Artificial Intelligence,}
  \institution{Fudan University}
  \city{Shanghai}
  \country{China}}
\email{zhaoyc22@m.fudan.edu.cn}

\author{Shisong Chen}
\affiliation{%
  \department{Shanghai Institute of Artificial Intelligence for Education,}
  \institution{East China Normal University}
  \city{Shanghai}
  \country{China}
  }
\email{sschen@stu.ecnu.edu.cn}

\author{Jiacheng Zhang}
\affiliation{%
  \department{College of Computer Science and Artificial Intelligence,}
  \institution{Fudan University}
  \city{Shanghai}
  \country{China}
}
\email{jiachengzhang22@m.fudan.edu.cn}

\author{Zhixu Li}
\authornote{Corresponding author.}
\affiliation{%
  \institution{School of Information, Renmin University of China; School of Smart Governance, Renmin University of China}
  \city{Beijing}
  \country{China}
}
\email{zhixuli@ruc.edu.cn}


\begin{abstract}
Recent advances in large language models (LLMs) have demonstrated impressive capabilities in code-related tasks such as code generation and automated program repair. Despite their promising performance, most existing approaches for code repair suffer from high training costs or computationally expensive inference. Retrieval-augmented generation (RAG), with its efficient in-context learning paradigm, offers a more scalable alternative. However, conventional retrieval strategies, which are often based on holistic code-text embeddings, fail to capture the structural intricacies of code, resulting in suboptimal retrieval quality. To address the above limitations, we propose \textbf{ReCode}, a fine-grained retrieval-augmented in-context learning framework designed for accurate and efficient code repair. Specifically, ReCode introduces two key innovations: (1) an algorithm-aware retrieval strategy that narrows the search space using preliminary algorithm type predictions; and (2) a modular dual-encoder architecture that separately processes code and textual inputs, enabling fine-grained semantic matching between input and retrieved contexts. Furthermore, we propose RACodeBench, a new benchmark constructed from real-world user-submitted buggy code, which addresses the limitations of synthetic benchmarks and supports realistic evaluation. Experimental results on RACodeBench and competitive programming datasets demonstrate that ReCode achieves higher repair accuracy with significantly reduced inference cost, highlighting its practical value for real-world code repair scenarios.
\end{abstract}


\maketitle

\section{Introduction}
Large language models have achieved significant breakthroughs in recent years, driven by both the continuous scaling of model parameters and advancements in training methodologies. State-of-the-art models such as GPT-4\cite{achiam2023gpt} and DeepSeek-R1\cite{guo2025deepseek} have demonstrated exceptional performance across diverse language understanding and generation tasks, while simultaneously catalyzing advancements in numerous downstream applications, which highlights their remarkable generalization capabilities. As LLM capabilities advance, research in code intelligence has increasingly focused on the field of code generation, where models synthesize executable program code from natural language specifications or contextual prompts. The development of Codex\cite{chen2021evaluating} marked a pivotal milestone in this direction, ultimately leading to practical deployments like GitHub Copilot and ushering in a new era of intelligent programming assistants. Subsequent releases of open-source models, including CodeLlama\cite{roziere2023code} and StarCoder\cite{li2023starcoder}, have further validated the robust capabilities of LLMs in comprehending code semantics, performing logical reasoning, and handling complex programming tasks with increasing sophistication.

\begin{figure}[h]
  \centering
  \includegraphics[width=0.9\linewidth]{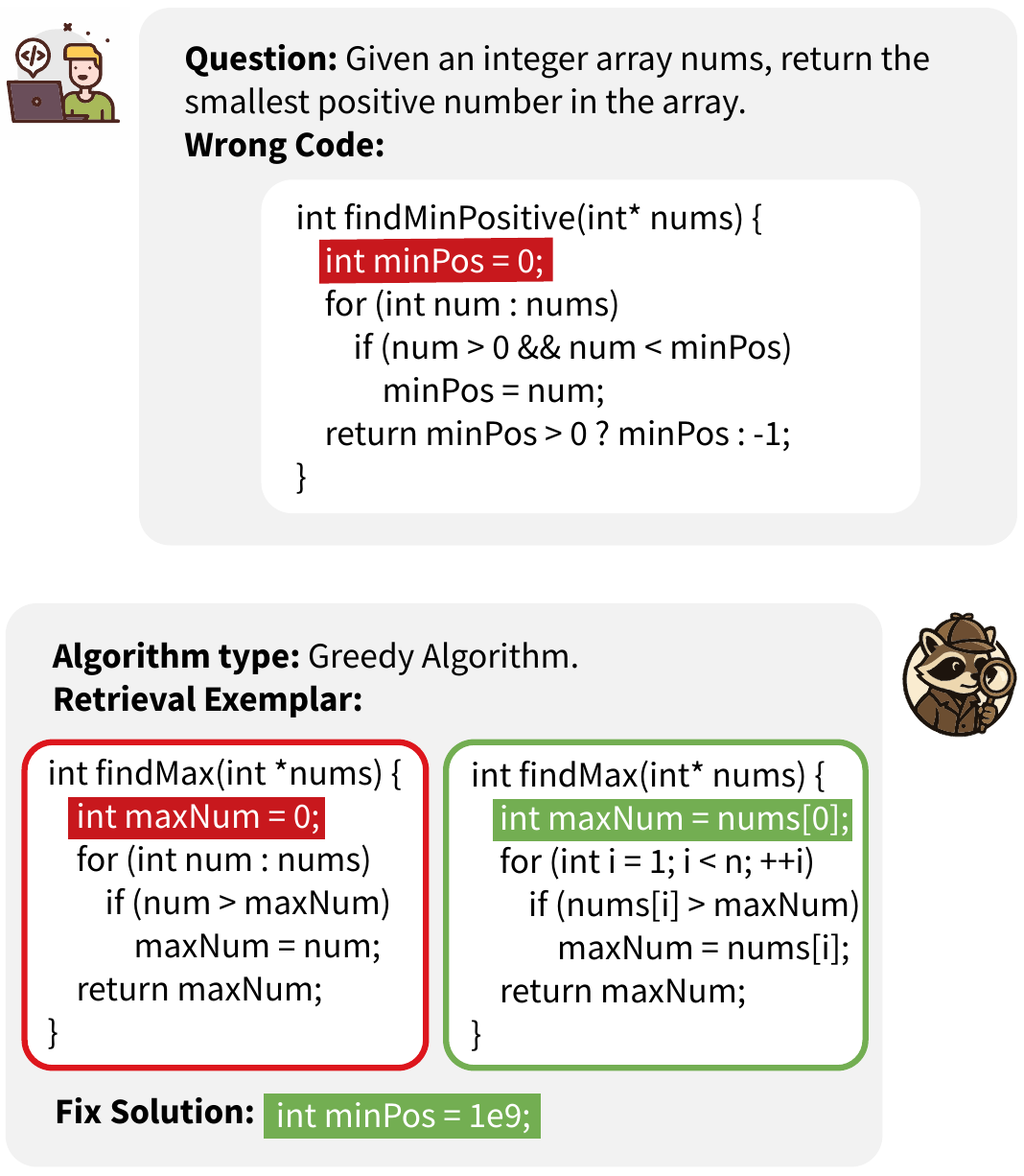}
  \caption{An example of our proposed ReCode Method. For a given question, ReCode first determines the algorithm of the wrong code to narrow the subset of our constructed code knowledge base. Then, the exemplar retrieved through our dual-view fused feature is connected with the user's query to prompt the LLM to give final solutions.}
\label{fig:example}
\end{figure}

Building on the remarkable capabilities of LLMs in code understanding and generation, recent research increasingly explores their potential in automated code repair, where models directly propose fixes for buggy programs. 
However, given the complex semantics and highly rigid syntax of programming languages, accurate code repair remains a non-trivial challenge for existing LLMs.
Many studies leverage inference-time strategies to mitigate current limitations. Sampling-based approaches, such as best-of-$N$\cite{liu2020learning} and self-consistency \cite{wang2022self}, improve repair quality by producing diverse candidates and selecting the most promising output. More recently, iterative self-repair methods have gained traction: instead of regenerating code from scratch, models refine initial predictions through multi-turn prompting and execution feedback \cite{chen2023teaching,le2022coderl,zhang2023self}, significantly enhancing efficiency while maintaining high repair accuracy \cite{olausson2023self}. In parallel, a complementary line of work employs supervised fine-tuning on curated repair datasets \cite{gupta2020synthesize,fu2019coda,chen2023improving}, enabling task-specific adaptation through additional training and labeled data.

Despite their effort, two key limitations consistently impede the practical deployment of previous code repair methods. Firstly, both training-based and inference-time approaches suffer from substantial computational costs: the former requires expensive human annotations and training, while the latter often incurs high inference-time latency due to multi-turn reasoning or sampling procedures. In addition, these methods exhibit limited adaptability to out-of-distribution (OOD) defects and novel repair patterns. Training-based methods struggle to incorporate emerging knowledge without frequent continued training, which is not only resource-intensive but also susceptible to catastrophic forgetting \cite{luo2023empirical}, which compromises performance on previously seen tasks. Meanwhile, inference-time strategies, though avoiding repeated retraining, are fundamentally trapped in the static knowledge gained during the training phase, making it difficult to generalize to unseen or structurally unfamiliar bug scenarios.

Motivated by the aforementioned limitations, we explore retrieval-augmented generation (RAG) as a promising alternative paradigm for enabling code repair without requiring parameter updates or additional training. However, our experiment results in Table~\ref{tab:valid} reveal the limitations of existing RAG methods in the context of code repair. Specifically, conventional RAG approaches typically encode problem descriptions and source code in a monolithic fashion. This unified representation neglects the inherent structure and semantics of code, thereby hindering the retrieval of relevant exemplars.

In this paper, we propose \textbf{ReCode}, a fine-grained retrieval-augmented generation framework that constructs modular representations for both source code and its accompanying textual descriptions. This design enables the model to capture domain-specific semantics effectively.

Beyond modular modeling, ReCode further leverages the model's intrinsic capacities in code comprehension and reasoning by introducing an algorithm-aware hybrid retrieval mechanism. 

Specifically, ReCode first uses the LLM to analyze the user's query to infer the algorithm type, which is then utilized to retrieve the most relevant exemplars from our pre-constructed algorithm-specific knowledge corpus. By integrating algorithmic categorization with the model’s semantic modeling towards the defective code, ReCode significantly enhances the contextual relevance of retrieved in-context exemplars and supports more accurate and adaptive code repair generation.

To further support rigorous and realistic evaluation, we construct a high-quality benchmark, \textbf{RACodeBench}, consisting of a systematic collection of real-world buggy–fixed code pairs manually curated and annotated from user submissions. RACodeBench reflects authentic software development scenarios and enables precise performance assessment. Extensive experiments conducted on both RACodeBench and several widely used competitive programming benchmarks demonstrate the superior repair accuracy and efficiency of our proposed ReCode.

Our contributions are summarized as follows:
\begin{itemize}
  \item We propose ReCode, a retrieval-augmented generation method that integrates algorithm-aware categorization and modular modeling to enable effective code repair.
  \item We curate RACodeBench, a benchmark centered on real-world user-submitted bugs, consisting of a large collection of authentic buggy–fixed code pairs, to support realistic and rigorous evaluation of code repair methods.
  \item We conduct systematic and comprehensive experiments, demonstrating that our proposed ReCode not only maintains strong repair performance but also significantly reduces inference cost.
\end{itemize}

\section{Related Work}
In this section, we give a brief introduction to the current research status of automatic program repair and in-context learning.

\subsection{Automatic Program Repair}
In addition to direct code generation, increasing attention has been given to automatic program repair (APR), which seeks to detect and fix code bugs automatically. Early APR approaches relied on template-based and search-based strategies \cite{ghanbari2019practical,peng2024domain,jiang2018shaping,yang2022transplantfix}, but their dependence on handcrafted rules limited generalization and scalability.

With the advent of deep learning, APR has witnessed a paradigm shift toward learning-based methods that model the repair process as a translation task from buggy code to its corrected version. Sequence-to-sequence models \cite{zhong2022neural, zhang2023survey, huang2023survey} trained on synthetic buggy–fixed pairs \cite{hu2022fix,chakraborty2020codit,jiang2021cure} improved generalization across bug types and languages, though the artificial–real bug gap remained a key challenge. Recently, large language models have shown strong zero- and few-shot capabilities in code repair \cite{chen2021evaluating,chowdhery2023palm,li2022competition,fried2022incoder}, offering new opportunities beyond supervised data.

Building on this, a growing body of research explores the ability of LLMs to repair their own generated code, marking a key direction in self-improving AI systems. \citet{wang2022self2} and \citet{burns2023weak} propose self-improvement mechanisms that leverage test case execution or natural language feedback to iteratively refine model outputs. More recent advancements have introduced structured frameworks for integrating feedback and reasoning into the repair process. \citet{zhang2023self} and \citet{olausson2023self} utilize execution traces and error messages to guide the model in understanding and correcting failures. \citet{shinn2023reflexion} extends this idea by using memory-augmented reasoning loops, allowing the model to reflect on previous attempts and improve over time. \citet{zhong2024debug} introduces a program decomposition strategy, breaking programs into basic blocks and monitoring runtime variable values after each block to verify semantic correctness with respect to the original specification. Similarly, \cite{shi2024code} adopts a hierarchical repair framework that separates the debugging process into low-level error localization and high-level semantic repair, significantly improving repair precision and interpretability.

Distinct from prior approaches, our method harnesses the in-context learning ability of large language models, enabling efficient and flexible code repair without model fine-tuning or reliance on auxiliary components.

\subsection{In-context Learning}

Despite the effectiveness of supervised fine-tuning across various tasks, it has several notable drawbacks. It typically requires access to a large volume of high-quality annotated data, which may be expensive or difficult to obtain. Additionally, the fine-tuning process is computationally intensive and can lead to undesirable side effects, such as catastrophic forgetting\cite{luo2023empirical}. To address these challenges, in-context learning has emerged as a promising alternative paradigm\cite{brown2020language,akyurek2022learning,von2023transformers}. In-context Learning enables LLMs to adapt to new tasks by conditioning on a small number of demonstration examples provided at inference time, without any modification to model parameters. This makes it a flexible and cost-effective approach to task generalization.

Recent studies explore the utility of in-context learning for code repair. For instance, \citet{yin2024thinkrepair} proposes an automated code self-correction framework that combines few-shot prompting with chain-of-thought (CoT) reasoning and execution feedback to iteratively guide the repair process. \citet{agarwal2024many} also leverage CoT rationales to enhance repair performance but intentionally omit these rationales from the input context during few-shot prompting in order to separate reasoning from learning signals. In another approach, \citet{mavalankar2025aupair} exploits the ICL capabilities of large language models and employs a filtering mechanism to select a small set of optimal repair examples generated by the model itself as in-context references. In contrast, our approach selects the most relevant and authentic examples corresponding to the current problem, rather than relying on model-generated repair instances.

\begin{figure*}[h]
  \centering
  \includegraphics[width=0.75\linewidth]{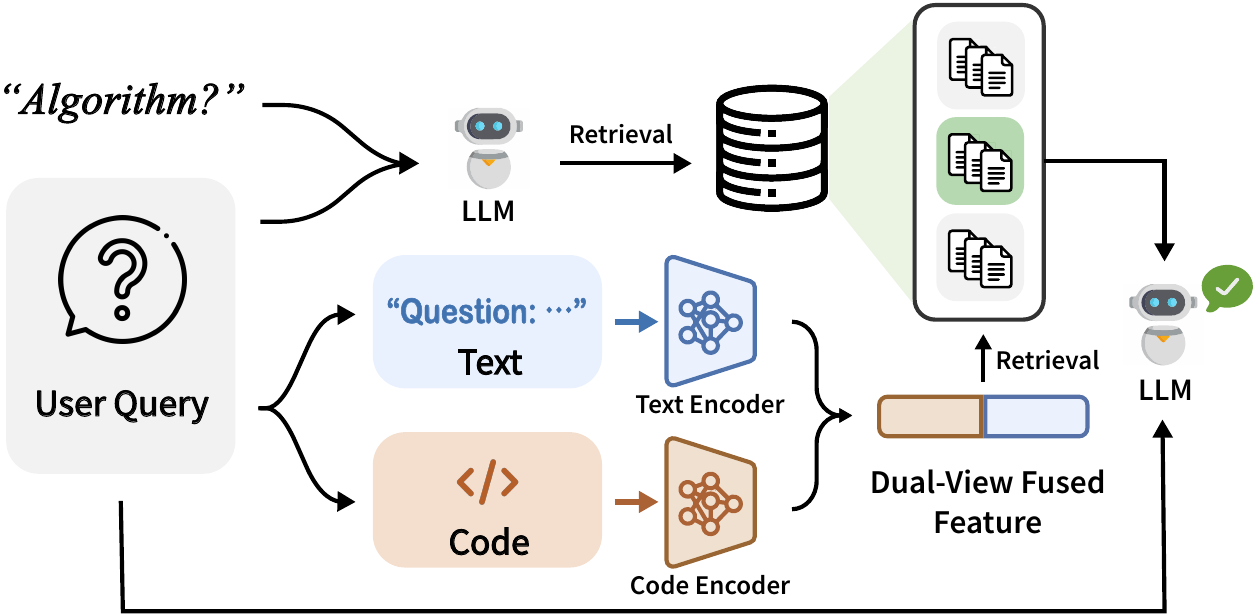}
  \caption{Overview of our proposed ReCode Method. Given a user query, the model predicts multiple algorithm type labels related to the buggy code. A dual encoding strategy extracts textual and code features to form a fused representation. Using predicted labels, semantic retrieval is performed in parallel across sub-knowledge bases. Retrieved results are integrated via a collaborative mechanism to build a semantically relevant in-context example set, which guides the model to generate accurate code repairs.}
\label{fig:框架图}
\end{figure*}

\section{Method}
In this section, we first provide the problem formulation and conduct experiments to demonstrate the existing limitations of traditional retrieval-augmented methods. Then, we meticulously introduce the framework of our ReCode. Finally, we elaborate on the construction of our knowledge base and RACodeBench.

\subsection{Problem Formulation}
The code repair task is formally defined as a conditional generation problem. Given a problem description \( q \in \mathcal{Q} \) and its corresponding defective code implementation \( x \in \mathcal{X} \), which may contain syntactic or semantic errors, the objective is to learn a generative model \( f_\theta \) capable of producing a corrected code \( y = f_\theta(q, x) \). The generated code must satisfy two fundamental requirements: it needs to be syntactically well-formed by strictly conforming to the grammar rules of the target programming language, and semantically correct by accurately implementing the intended functionality specified in \( q \), which is formally verified through a comprehensive set of test cases \( \mathcal{T}_q \) that serve as executable specifications for correctness.

Within the retrieval-augmented in-context learning framework, an external support set \( \mathcal{D} = \{(q_i, x_i, y_i)\}_{i=1}^N \) is introduced to provide relevant auxiliary information. For a target input \((q, x)\), a retrieval function
\( \mathcal{R}(q, x; \mathcal{D}) \) identifies the most relevant example from the support set \(\mathcal{D}\):
\begin{equation}
(q^*, x^*, y^*) = \mathcal{R}(q, x; \mathcal{D})
\end{equation}

Subsequently, this retrieved example is concatenated with the target input to form the contextual prompt of the model:
\begin{equation}
\text{Prompt} = \operatorname{Concat}\big((q^*, x^*, y^*), (q, x)\big)
\end{equation}

The model then generates the repaired code based on the constructed prompt:
\begin{equation}
y = f_\theta(\text{Prompt}) \quad \text{such that} \quad y \models \mathcal{T}_q
\end{equation}

Here \(y \models \mathcal{T}_q\) denotes that the generated code \(y\) satisfies all test cases in \(\mathcal{T}_q\), thus ensuring both syntactic validity and semantic correctness.

A central challenge of this approach lies in the design of the retrieval function \( \mathcal{R} \), which must effectively identify examples that are both semantically informative and structurally similar to the target input, thus providing valuable guidance for the generation process.

\subsection{Limitations of Traditional Retrieval-Augmented Methods}

Although retrieval-augmented in-context learning frameworks have been empirically validated as effective in code repair tasks, conventional retrieval mechanisms within these frameworks often rely on a unified encoding strategy. This approach encodes the problem description and the associated buggy code as a single input, retrieving support examples based on the overall similarity of their combined representations. However, this retrieval paradigm exhibits several notable limitations: it inadequately accounts for structural variations in code, fails to capture fine-grained error semantics, and often lacks precision in identifying the intended repair operations. These deficiencies reduce the contextual relevance and corrective utility of the retrieved exemplars, ultimately limiting the effectiveness of the repair process. As shown in Table~\ref{tab:valid}, replacing the unified encoding scheme with our proposed dual encoding strategy, which independently models textual and code representations, leads to consistent improvements in test pass rates on RACodeBench across both the Gemma-2 \cite{team2024gemma} and GPT-4o-mini.

\begin{table}[ht]
\centering
\caption{Test Pass Rates of Unified vs. Dual Encoding Strategies on RACodeBench}
\begin{tabular}{lc}
\toprule
\textbf{Encoding Strategy} & \textbf{Test Pass Rate} \\
\midrule
Gemma-2-27B + Unified Encoding     & 26.25\%                  \\
Gemma-2-27B + Dual Encoding        & \textbf{27.73}\%                  \\
GPT-4o-mini + Unified Encoding     & 36.72\%                  \\
GPT-4o-mini + Dual Encoding       & \textbf{38.49}\%                  \\
\bottomrule
\end{tabular}
\label{tab:valid} 
\end{table}

Apart from encoding inefficiencies, existing traditional retrieval methods often overlook the model’s inherent ability to understand the input code. Instead, they rely solely on surface-level similarity, which can lead to mismatches between the retrieved examples and the actual repair needs. This lack of semantic and structural alignment reduces the effectiveness of the retrieved examples and limits the overall performance of the repair process.

\subsection{ReCode Framework}

To overcome the limitations inherent in traditional retrieval-augmented approaches, we propose \textbf{ReCode}, a novel algorithm-aware framework that integrates dual-view encoding with hybrid retrieval mechanisms to enhance the precision and contextual relevance of retrieved examples for code repair.

An overview of the ReCode framework is illustrated in Figure~\ref{fig:框架图}. The overall process comprises two core modules: 

\paragraph{\textbf{Algorithm-Aware}}

The algorithm-aware module infers the underlying algorithmic intent of a buggy code snippet by leveraging the advanced reasoning and comprehension capabilities of LLMs to analyze code structure and semantics beyond surface-level patterns. Recognizing that real-world code frequently integrates multiple algorithmic paradigms, such as graph traversal combined with greedy heuristics, this module adopts a multi-label classification approach, assigning multiple algorithm categories to each snippet. This strategy effectively captures a broad spectrum of potential error types associated with diverse algorithms. By exploiting the intrinsic understanding of LLMs, the module achieves a more precise characterization of algorithmic signals, thereby improving the relevance and efficacy of retrieved repair exemplars.

\paragraph{\textbf{Dual-View Encoding}}

The dual-view encoding module employs specialized encoders to independently process the two input modalities. Specifically, the natural language description is encoded via a dedicated text encoder designed to capture task-level semantic intent, while the buggy code snippet is processed by a code encoder tailored to preserve the syntactic and structural features intrinsic to source code. Following this modality-specific encoding, a feature fusion step integrates the textual and code representations into a unified vector. This fusion synergistically combines complementary information from both views while maintaining their distinctive properties. In contrast to conventional unified encoding methods, which concatenate text and code into a single sequence and often result in modal interference and dilution of critical semantic or structural cues, our approach preserves the unique characteristics of each modality. This separation enables more precise modeling of the inherent features of natural language and source code, yielding fused representations that are both richer and more discriminative. Such enhanced expressiveness and modality-specific focus are advantageous for the subsequent retrieval phase.

Given a user query, ReCode first decomposes the input into two distinct components: a natural language description and a buggy code snippet. These two modalities are then independently processed through the dual-view encoding module to generate rich and discriminative representations. Leveraging the algorithm-aware module, ReCode predicts multiple algorithmic categories relevant to the buggy code, which guides the retrieval process. Based on these predictions, the hybrid retrieval module selectively searches corresponding algorithm-specific sub-knowledge bases to retrieve the most relevant repair exemplars. The retrieved examples are then aggregated and ranked to form a final set of contextually aligned support instances. Finally, these carefully selected examples are provided as guidance to downstream models, enabling more accurate and effective code repair.

\subsection{Construction of RACodeBench and Knowledge Bases}
\label{sec:benchmark}
To support our proposed ReCode method, we construct two complementary resources derived from the widely adopted competitive programming platform Codeforces: a hierarchically structured knowledge base and a standardized evaluation benchmark, RACodeBench. Both resources are built from large-scale collections of real-world programming problems and user submissions, ensuring representativeness and diversity across algorithmic domains.

\paragraph{\textbf{Knowledge Base Construction}}
We build the knowledge base from the historical archives of Codeforces, leveraging both problem statements and associated user submissions. Using the original algorithmic tags (e.g., greedy, binary search, graph, and dp), we organize the knowledge base into a multi-level hierarchy. At the top are broad algorithm categories, each containing numerous problems. Under each problem, we curate paired samples of erroneous and corrected submissions.

The erroneous submissions capture common syntactic and semantic faults, while the corrected counterparts represent valid solutions that pass all test cases. To enable fine-grained error retrieval and analysis, we employ a semi-automated pipeline that performs differential analysis between buggy and fixed submissions, enriched with compiler diagnostics and execution outcomes. Each pair is annotated with detailed error types, enabling precise access to error-specific and algorithm-specific examples within the knowledge base.

\paragraph{\textbf{RACodeBench Construction}}  
To comprehensively evaluate code repair models, we construct RACodeBench, a large-scale, diagnostic benchmark built upon the historical archives of Codeforces. RACodeBench is specifically designed to capture a wide range of real-world programming errors and algorithmic challenges, enabling fine-grained and algorithm-aware assessment of model performance. Each instance in RACodeBench consists of the following elements:
(1) the natural language description of a programming problem,
(2) an erroneous user-submitted code that fails due to compilation or runtime/test-case errors,
(3) the corrected version of the submission that passes all test cases,
(4) the full set of public test cases provided by Codeforces,
(5) a fine-grained annotation of the error type (e.g., syntax error, incorrect loop boundary), and
(6) a code-level diff highlighting the exact changes between the buggy and repaired versions.

To enable comprehensive analysis and systematic comparison, all error annotations are generated through a meticulously designed semi-automated pipeline. This process involves: (1) conducting syntactic and semantic differencing between erroneous and corrected code segments, (2) integrating compiler diagnostics and test execution outputs, and (3) performing manual verification on a carefully selected representative subset to guarantee annotation quality and consistency. These multi-faceted annotations transform the benchmark into not merely a quantitative metric for test pass rates, but more importantly, a sophisticated diagnostic tool for fine-grained analysis of model behavior. During dataset construction, we implement rigorous measures to ensure diversity and balance across multiple dimensions. The problem set encompasses a comprehensive spectrum of algorithmic domains, including but not limited to dynamic programming, greedy algorithms, graph traversal, and number theory. Problem difficulty is systematically stratified according to Codeforces ratings, ensuring balanced evaluation across both fundamental programming concepts and advanced algorithmic challenges. Furthermore, we employ stratified sampling of error types to mitigate distributional bias, thereby preventing any single repair category from disproportionately influencing the benchmark results.

To maintain experimental rigor and preclude potential data contamination, RACodeBench implements strict partitioning between the benchmark dataset and the retrieval knowledge base. This partitioning guarantees that no problem instances or user submissions appearing in the evaluation set are included in the in-context learning knowledge base. Such stringent separation preserves the integrity of performance assessment by simulating authentic code repair scenarios, where models must address genuinely novel problems. This design principle ensures that evaluation results accurately reflect model capabilities in real-world application contexts.

\section{Experiments}

\begin{table*}[ht]
\centering
\begin{tabular}{c|c|ccc|ccc}
\toprule
 & \multirow{2}{*}{\textbf{Model}} & \multicolumn{3}{c|}{\textbf{Test Pass Rate(\%)}} & \multicolumn{3}{c}{\textbf{Strict Accuracy(\%)}} \\
\cline{3-8}  
                &                & Best-of-$N$ & Self-repair & Ours & Best-of-$N$ & Self-repair & Ours \\
\hline  
\hline
\multirow{4}{*}{Open-source}           & Gemma-2-9B & 23.34 & 22.67 & \textbf{26.07} & 15.83 & 15.00 & \textbf{16.67} \\
                                       & Gemma-2-27B & 24.96 & 22.74 & \textbf{29.83} & 17.08 & 15.83 & \textbf{19.17} \\
                                       & DeepSeek-V2-Chat & 30.28 & 29.32 & \textbf{38.04} & 19.17 & 22.08 & \textbf{26.25} \\
                                       & DeepSeek-Coder-V2-Instruct & 31.54 & 33.08 & \textbf{40.47} & 20.42 & 23.17 & \textbf{28.75} \\
\hline
\multirow{2}{*}{Closed-source}         & Gemini-1.5-Flash & 28.88 & 33.90 & \textbf{37.96} & 19.17 & 23.75 & \textbf{26.67} \\
                                       & GPT-4o-mini & 31.09 & 34.79 & \textbf{41.06} & 21.25 & 24.58 & \textbf{30.41} \\
\bottomrule
\end{tabular}

\caption{Code repair performance on RACodeBench. With $N=8$ inference calls, we evaluate best-of-$N$, self-repair, and ReCode using test pass rate and strict accuracy metrics. Across all tested models, ReCode consistently achieves superior performance.}
\label{tab:main-results}
\end{table*}

\subsection{Experimental Setup}

\textbf{Datasets:} We perform in-distribution evaluation on our newly constructed dataset, \textbf{RACodeBench}, which contains buggy code samples meticulously annotated with corresponding algorithmic error types. To further assess the out-of-distribution generalization capability of our method, we conduct evaluations on six additional competitive programming datasets: (1) AtCoder (800 problems), (2) CodeChef (2.4k problems), (3) HackerRank (720 problems), (4) HackerEarth (1k problems), (5) GeeksforGeeks (680 problems), and (6) Aizu (1.8k problems). These benchmarks collectively cover a wide spectrum of problem domains, programming languages, and difficulty levels, providing a rigorous and comprehensive testbed for evaluating the robustness and generalization of code repair models.

\textbf{Models:} For retrieval, we use OASIS-code-1.3B\cite{gao2025oasis} as the code encoder and bge-m3\cite{chen2024bge} as the text encoder. For inference, we evaluate our approach using a set of open-source and closed-source LLMs, including Gemma-2-9B, Gemma-2-27B\cite{team2024gemma}, GPT-4o-mini, Gemini-1.5-Flash, DeepSeek-V2-Chat\cite{liu2024deepseek}, and DeepSeek-Coder-V2-Instruct\cite{zhu2024deepseek}.

\textbf{Metrics:} The most direct and reliable way to evaluate code repair performance is by executing the generated code against test cases. We adopt and extend two commonly used metrics—test pass rate and strict accuracy \cite{hendrycks2021measuring}—to provide a comprehensive assessment of model effectiveness. To mitigate variability in model outputs, we consider the best result among the $N$ candidate solutions generated by the language models.

Formally, let $P$ denote the total number of problems. For each problem $p$, let $\text{code}p^j$ represent the $j$-th generated candidate solution. Each problem is associated with a set of $K$ test cases $\mathcal{T}p = {(x{p,1}, y{p,1}), \ldots, (x_{p,K}, y_{p,K})}$, where $x_{p,k}$ is the input and $y_{p,k}$ is the expected output.

The \textbf{test pass rate} is defined as the average proportion of test cases passed across all problems:
\begin{equation}
\frac{1}{P} \sum_{p=1}^{P} \max_{1 \le j \le N}  \frac{1}{{|K|}} \sum_{k=1}^{|K|} \mathbf{1} \left\{ \text{eval}(\text{code}_p^j, x_{p,k}) == y_{p,k} \right\}
\end{equation}

Here, the inner summation computes the fraction of test cases passed by the $j$-th candidate for problem $p$, and the outer summation averages this score over all problems. The $\max$ operator selects the best-performing candidate among the $N$ generated solutions.

The \textbf{strict accuracy} metric grants credit only if a candidate passes all test cases for a given problem:
\begin{equation}
\frac{1}{P} \sum_{p=1}^{P} \max_{1 \le j \le N} \prod_{k=1}^{|K|} \mathbf{1} \left\{ \text{eval}(\text{code}_p^j, x_{p,k}) == y_{p,k} \right\}
\end{equation}

In this expression, the product equals 1 only if the candidate passes every test case for problem $p$, and 0 otherwise. Similar to the test pass rate, the maximum over candidates is taken to identify the best attempt, and the average is computed over all problems.

\textbf{Baselines:}
We compare ReCode with two representative baselines: best-of-$N$\cite{liu2020learning} and self-repair \cite{olausson2023self}. The best-of-$N$ leverages multiple LLM calls at inference time to enhance performance by generating $N$ diverse candidate completions and selecting the one that achieves the best downstream evaluation result. To promote diversity among the generated candidates, we set the sampling temperature to 1.0, following established practice~\cite{renze2024effect}. The self-repair adopts a multi-stage strategy in which the total LLM calls at inference time are distributed across a sequence of generation and refinement steps. Specifically, a portion of the LLM calls is allocated to generating natural language feedback that identifies potential issues in the initial code, while the remaining calls are used to produce revised code conditioned on this feedback. In ReCode, the inference-time budget is similarly divided, with part of the calls used for algorithm classification and the rest for retrieval-augmented code generation. We adopt different values of $N$ depending on the evaluation setting. For in-distribution experiments, where buggy code is available, we set $N = 8$. For out-of-distribution experiments, where the model must first synthesize an initial solution before performing repair, we increase $N$ to 32 to account for the additional generation complexity.

\subsection{Main Results}

\begin{figure*}[h]
  \centering
  \includegraphics[width=0.90\linewidth]{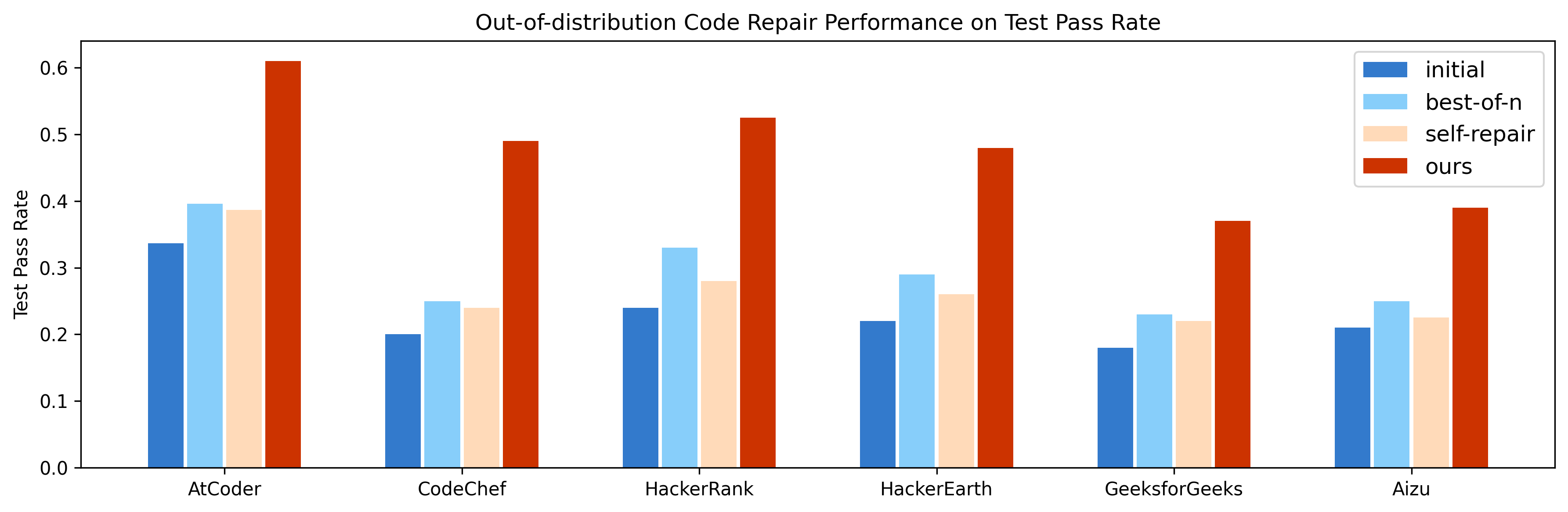}
  \caption{Test pass rates on out-of-distribution code repair. With $N = 32$ LLM calls at inference time, results using GPT-4o-mini show significant improvements across several common out-of-distribution competitive programming datasets, with our approach consistently outperforming all baseline methods, demonstrating its strong generalization capability.}
\label{figure-3}
\end{figure*}

\begin{figure*}[h]
  \centering
  \includegraphics[width=0.90\linewidth]{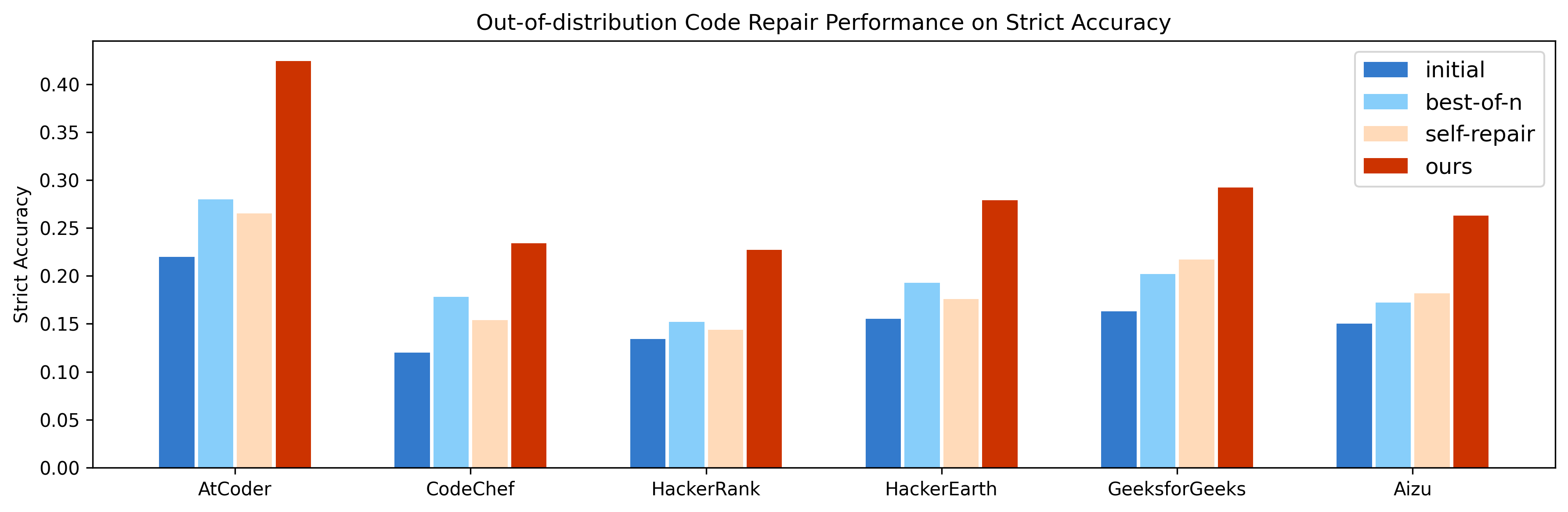}
  \caption{Strict accuracy on out-of-distribution code repair. With $N = 32$ LLM calls at inference time, results using GPT-4o-mini demonstrate that our approach strongly solves competitive programming problems, consistently outperforming all baseline methods.}
  \label{figure-4}
\end{figure*}
We report the core experimental results across both in-distribution and out-of-distribution settings. Detailed experimental results are summarized in Table~\ref{tab:main-results} for the in-distribution setting, and illustrated in Figure~\ref{figure-3} and Figure~\ref{figure-4} for the out-of-distribution evaluation.

As shown in Table~\ref{tab:main-results}, the in-distribution experimental results on RACodeBench demonstrate that our method consistently outperforms competing approaches across all models and evaluation metrics. This superior performance stems from several key factors intrinsic to our approach. First, by effectively integrating external, domain-specific knowledge from specific algorithmic domains through a retrieval mechanism, our method provides precise and contextually relevant guidance that purely generative methods relying solely on implicit model knowledge often fail to capture. Second, leveraging a hierarchically structured knowledge base organized by algorithmic categories allows the model to access tailored exemplars that closely match the semantic and structural patterns of the target problem, enabling more targeted and semantically informed code repairs. Finally, the combination of retrieval and generation creates a complementary synergy where external knowledge enriches the model’s internal representations, leading to enhanced generalization and robustness in correcting a diverse spectrum of complex programming errors.

Notably, when scaling from Gemma-2-9B to Gemma-2-27B, baseline methods show only modest improvements: Best-of-$N$ increases by 6.94\% (from 23.34 to 24.96), and Self-repair by just 0.3\% (from 22.67 to 22.74).  In contrast, our retrieval-augmented approach improves significantly by 14.42\% (from 26.07 to 29.83).  These results suggest that while larger models do enhance performance, such improvements are relatively limited for methods that do not incorporate external context.  However, when augmented with retrieval, larger models, thanks to their enhanced in-context learning capacity, can better exploit external knowledge.  By retrieving targeted exemplars from a hierarchical knowledge base, our approach enables deeper contextual understanding. This demonstrates that the benefit of model scaling is amplified when coupled with retrieval-based augmentation.

Further analysis of the DeepSeek series reveals a clear trend: DeepSeek-V2-Chat, a general-purpose model, demonstrates a more substantial performance improvement from our approach than DeepSeek-Coder-V2, which is further pre-trained from an intermediate checkpoint of DeepSeek-V2 with code-specific objectives.  This indicates that models lacking explicit adaptation to code-related tasks benefit more from the external retrieval mechanism.  These findings suggest that our method is particularly effective for models with limited inherent code understanding, as the retrieved exemplars can compensate for their weaker internal representations and provide stronger task-specific guidance.

\begin{figure}[t]
  \centering
  \includegraphics[width=0.9\linewidth]{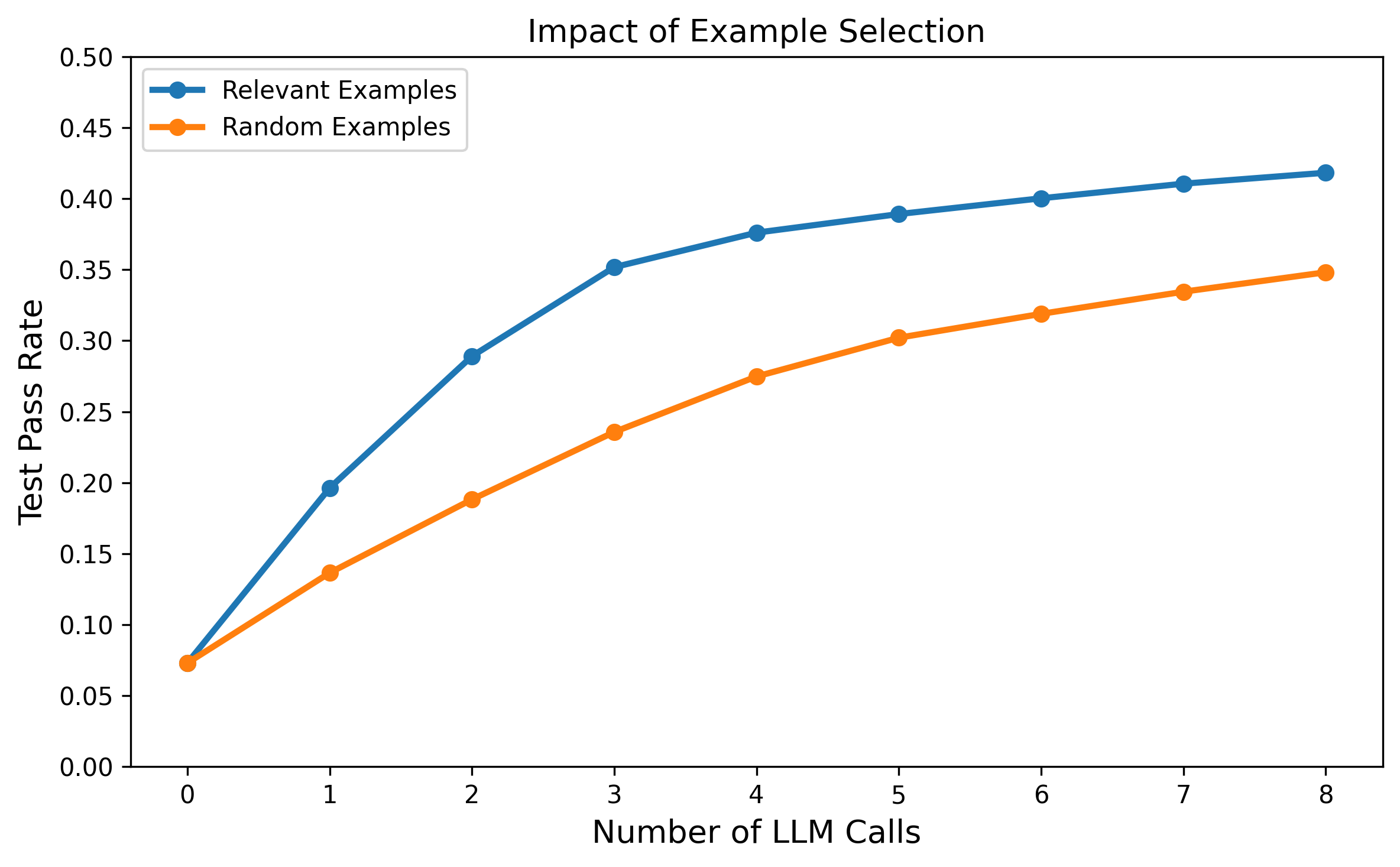}
  \caption{Impact of example selection. Selecting examples relevant to the current problem leads to consistently better performance under limited inference budgets (up to 8 LLM calls in this setting).}
\label{figure-5}
\end{figure}

Encouraged by the strong performance of our method in in-distribution experiments, we further evaluate its generalizability through out-of-distribution experiments. These experiments assess the efficacy of our approach by retrieving examples from Codeforces and testing on a distinct competition code dataset, thereby simulating real-world domain shifts. 

Figure~\ref{figure-3} and Figure~\ref{figure-4} present a comparative analysis of our proposed method against two baseline approaches, evaluated using the GPT-4o-mini model across test pass rate and strict accuracy metrics. The notable improvement in test pass rate validates the effectiveness of retrieving repair examples aligned with similar algorithmic categories. This targeted retrieval strategy significantly enhances the model’s ability to correct and improve initial solutions, even under out-of-distribution conditions. The advantage arises from algorithm classification-driven exemplar selection, which ensures provision of contextually and algorithmically relevant repair instances, thereby enabling robust generalization across varied datasets by leveraging common algorithmic patterns despite differences in problem specifics or data distribution. Likewise, our approach surpasses baselines in strict accuracy, reflecting the proportion of fully correct solutions passing all test cases, demonstrating its capability to generate complete and reliable solutions. This improvement stems from enriched contextual guidance that aids the model in addressing subtle errors and edge cases crucial for full correctness. Importantly, these gains highlight the dual benefits of our retrieval-augmented framework in enhancing both incremental repair quality and overall solution robustness, particularly in out-of-distribution scenarios.

\subsection{Impact of Example Selection}

We conducted an ablation study to rule out the possibility that performance gains in code repair tasks arise merely from the use of examples, regardless of their relevance. In the baseline setting, examples were randomly sampled from the entire pool. To ensure a fair comparison, the additional LLM call for algorithm classification was omitted in this setting. Experimental results(see Figure~\ref{figure-5}) demonstrate that our method, which retrieves examples highly relevant to the current problem, significantly outperforms the baseline. This performance gap underscores the importance of contextual and algorithmic alignment in example selection. Relevant examples often share underlying algorithmic patterns and problem-solving structures with the target problem, enabling the model to make more accurate and efficient repairs. In contrast, randomly sampled examples may better reflect the general task format but lack the specific structural cues necessary for precise correction.

Furthermore, under a limited LLM inference budget, the baseline exhibits only slow, marginal gains, whereas our method rapidly converges to high performance.  This reflects that retrieving highly relevant examples offers immediate, effective guidance, enabling efficient repair within fewer inference steps.  Such early convergence underscores the efficiency and efficacy of targeted retrieval in resource-constrained settings.

\begin{figure}[t]
  \centering
  \includegraphics[width=1\linewidth]{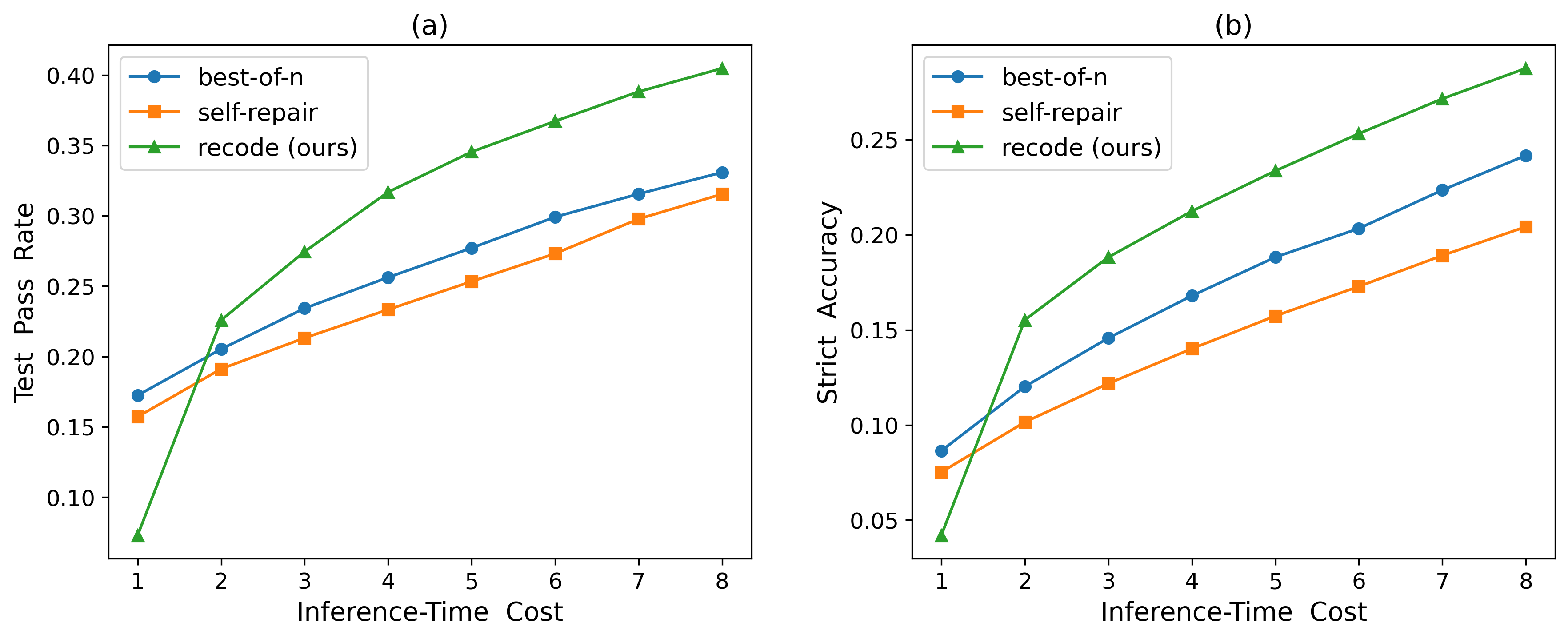}
  \caption{Comparative Analysis of Repair Performance and Inference Efficiency Across Methods Under In-Distribution Settings (using GPT-4o-mini). (a) Test Pass Rate metric.(b) Strict Accuracy metric. }
\label{figure-6}
\end{figure}

\begin{figure*}[h]
  \centering
  \includegraphics[width=0.85\linewidth]{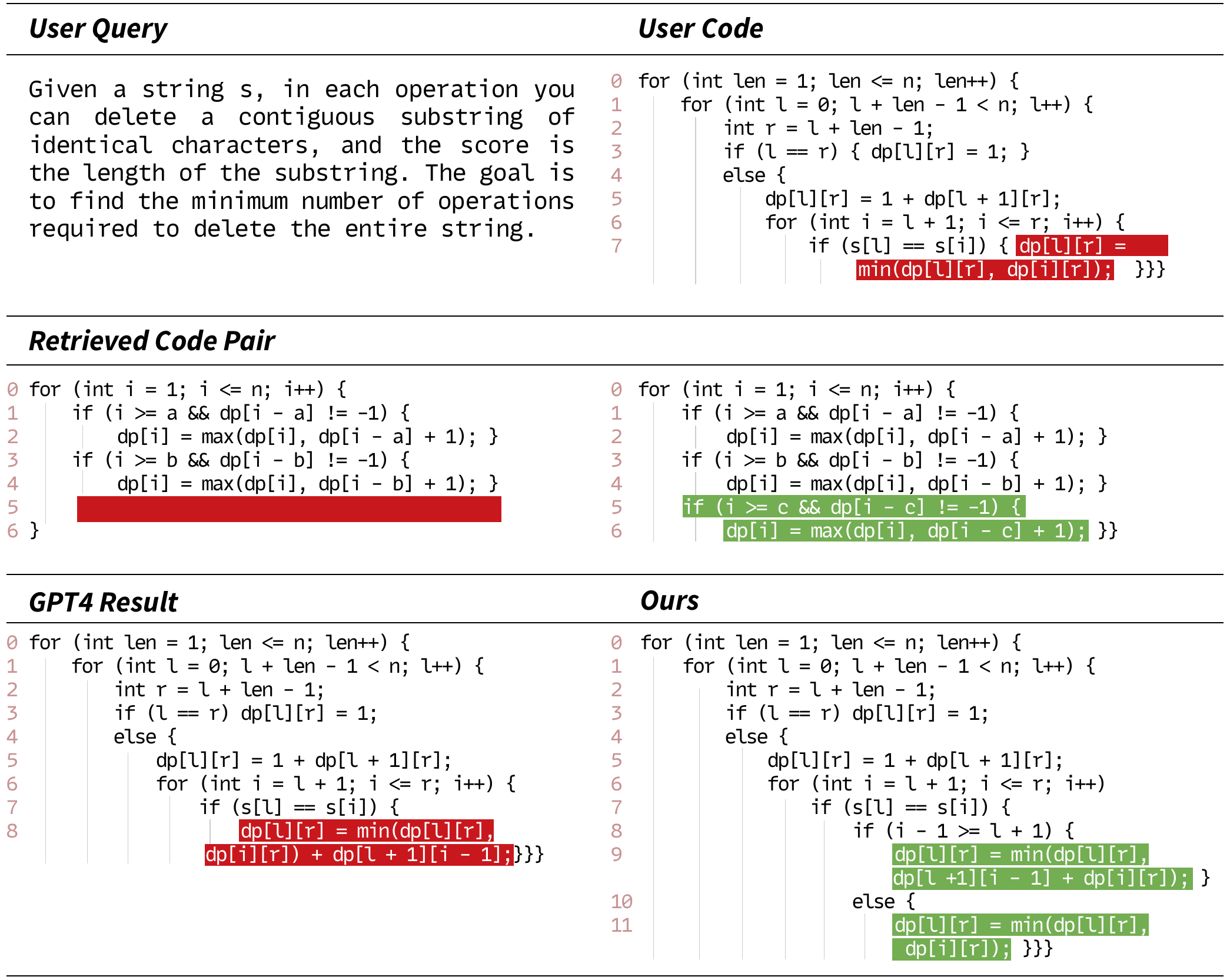}
  \caption{Qualitative Results of the ReCode method on our RACodeBench Benchmark.}
\label{fig:qualitity}
\end{figure*}

\subsection{Inference Cost Analysis}
In real-world automated code repair systems, inference efficiency is as critical as repair accuracy. High computational overhead increases response latency and limits system scalability and deployment. To address this, we systematically analyze the inference efficiency of Best-of-$N$, Self-Repair, and our proposed method, ReCode, under both in-distribution and out-of-distribution scenarios.

In the in-distribution experiments with a generation budget of $N=8$ calls, Best-of-$N$ independently samples 8 candidate fixes and selects the best, Self-Repair performs iterative corrections, and ReCode generates a high-quality repair in a single pass using context exemplars. Results show that ReCode performs slightly worse at the initial call ($N=1$) due to its initial algorithm type identification step. However, from $N=2$ onward, ReCode rapidly surpasses the others in test pass rate, effectively halving inference cost. At $N=4$, ReCode achieves comparable performance to others at $N=8$, effectively halving inference cost. Strict accuracy metrics confirm this trend, with ReCode reaching the upper bound of other methods by $N=5$, demonstrating efficient repair quality with reduced cost.

These performance differences arise because Self-Repair suffers from error accumulation during iterations, and Best-of-$N$ explores a large search space with many low-quality candidates, leading to inefficiency. In contrast, ReCode leverages carefully selected context exemplars to guide the model’s focus, enabling fast convergence to high-quality solutions and balancing accuracy with inference cost.

On the more challenging out-of-distribution AtCoder dataset, we measure the number of model calls needed to reach specific test pass rate thresholds. Starting from 33.72\%, ReCode requires significantly fewer calls—for example, only 4 calls to reach 35.0\%, whereas Best-of-$N$ and Self-Repair need 11 and 15 calls, respectively—reducing inference overhead by roughly 3 to 4 times. Furthermore, ReCode delivers larger performance gains per call, showcasing its efficient and targeted repair strategy.

In summary, ReCode achieves competitive or better repair accuracy while significantly lowering inference computation, making it more practical for real-world automated code repair applications.

\begin{table}[ht]
\centering
\caption{Inference Cost to Reach Test Pass Rate Thresholds on AtCoder, evaluated with GPT-4o-mini.}
\label{table-2}
\begin{tabular}{c|ccc}
\toprule
\multirow{2}{*}{\textbf{Test Pass Rate(\%)}} & \multicolumn{3}{c}{\textbf{Number of LLM Calls}} \\
\cmidrule(lr){2-4}
 & \textbf{Best-of-$N$} & \textbf{Self-Repair} & \textbf{ReCode} \\
\midrule
35.0 &  11 &  15 & \textbf{4} \\
35.5 &  12 &  15 & \textbf{4}\\
36.0 &  16 &  19 & \textbf{5}\\
36.5 &  16 &  21 & \textbf{5}\\
37.0 &  18 &  24 & \textbf{5}\\
37.5 &  21 &  27 & \textbf{6}\\
\bottomrule
\end{tabular}
\label{tab:xiaolv}
\end{table}

\subsection{Quality Result}
To further demonstrate our method’s effectiveness, Figure~\ref{fig:qualitity} presents a representative example. In this case, GPT-4o-mini attempts a direct repair but produces an incomplete solution. By contrast, ReCode retrieves a high-quality exemplar with a dynamic programming pattern and successfully transfers its underlying logic to the user’s code, resulting in a functionally complete repair. Moreover, ReCode not only ensures semantic correctness but also preserves the user’s original coding style and structure, underscoring the advantages of exemplar-guided generation.

\section{Conclusions}
We propose ReCode, a fine-grained retrieval-augmented generation framework that leverages modular code representations and an algorithm-aware retrieval strategy to enhance automated code repair.  ReCode improves the relevance of retrieved exemplars and enables accurate, efficient repair without additional training.  Extensive experiments on RACodeBench and competitive programming datasets show that ReCode outperforms existing methods in both repair accuracy and inference efficiency.  This work demonstrates the practical advantages of integrating retrieval with large language models for scalable and adaptive code repair.

\begin{acks}
This work is supported by the Key-Area Research and Development Program of Guangdong Province (2024B0101050005), Suzhou Key Laboratory of Artificial Intelligence and Social Governance Technologies (SZS2023007), Smart Social Governance Technology and Innovative Application Platform (YZCXPT2023101), and the Leadership Talent Program (Science and Education) of SIP.
\end{acks}

\section*{GenAI Usage Disclosure}
During the preparation of this work, the authors utilized GPT-4 for language refinement and sentence polishing. After using this tool, the authors reviewed and edited the content as needed and take full responsibility for the content of the publication.




\bibliographystyle{ACM-Reference-Format}
\bibliography{sample-base}

\appendix

\end{document}